\documentclass[final]{elsart}
\usepackage{ifpdf}
\usepackage{graphicx,natbib,amssymb,lineno}
\usepackage{bm}
\usepackage{amsmath}

\ifpdf
\usepackage[%
  pdftitle={},%
  pdfauthor={},%
  pdfsubject={},%
  pdfkeywords={},%
  pdfstartview=FitH,%
  bookmarks=true,%
  bookmarksopen=true,%
  breaklinks=true,%
  colorlinks=true,%
  linkcolor=blue,anchorcolor=blue,%
  citecolor=blue,filecolor=blue,%
  menucolor=blue,pagecolor=blue,%
  urlcolor=blue]{hyperref}
\else
\usepackage[%
  breaklinks=true,%
  colorlinks=true,%
  linkcolor=blue,anchorcolor=blue,%
  citecolor=blue,filecolor=blue,%
  menucolor=blue,pagecolor=blue,%
  urlcolor=blue]{hyperref}
\fi

\makeatletter
\def\elsartstyle{%
    \def\normalsize{\@setfontsize\normalsize\@xiipt{14.5}}
    \def\small{\@setfontsize\small\@xipt{13.6}}
    \let\footnotesize=\small
    \def\large{\@setfontsize\large\@xivpt{18}}
    \def\Large{\@setfontsize\Large\@xviipt{22}}
    \skip\@mpfootins = 18\p@ \@plus 2\p@
    \normalsize
}
\@ifundefined{square}{}{}
\makeatother

\begin{document}

\begin{frontmatter}
\title{Twisting Carbon Nanotubes: A Molecular Dynamics Study.}

\author{Zhao Wang}
\address{EMPA - Swiss Federal Laboratories for Materials Testing and Research, Feuerwerkerstrasse 39, CH-3602 Thun, Switzerland.}
\ead{wzzhao@yahoo.fr}
\ead[url]{wangwzhao.googlepages.com}

\author{Michel Devel, Bernard Dulmet}
\address{FEMTO-ST, ENSMM, 26 Chemin de l'\'{e}pitaphe, F-25030 Besan\c {c}on, France.}

\begin{abstract}
We simulate the twist of carbon nanotubes using atomic molecular dynamic simulations. The ultimate twist angle per unit length and the deformation energy are calculated for nanotubes of different geometries. It is found that the big tube is harder to be twisted while the small tube exhibits higher ultimate twisting ratio. For multi-walled nanotubes, the zigzag tube is found to be able to stand more deformation than the armchair one. We observed the surface transformation during twisting. Formation of structural defects is observed prior to fracture.  
\end{abstract}

\begin{keyword}
carbon nanotubes, molecular dynamics, twist, NEMS
\PACS {61.46.Fg, 81.40.Jj, 62.20.Dc, 31.15.Qg}
\end{keyword}
\end{frontmatter}

\section{Introduction}
\label{intro}
\linenumbers 
Carbon nanotube (CNT) is known as one of the strongest nanostructures currently known to mankind. This results from the high covalent energy of the conjugated bonds between quasi-$sp^{2}$ carbon atoms. Their Young's moduli is nearly 1 TPa and their ultimate stress can be up to 60 GPa \cite{Ruoff2003, Rafii-Tabar2004}. The thermal conductivity of CNTs is also very high (about 4000 W/m.K) \cite{Hone1999, Berber2000}. This makes them promising for future nanoelectromechanical systems (NEMS). Recently, it has been reported that CNTs can be used as key rotational elements in a nanoactuator \cite{Fennimore2003a} and in an electromechanical quantum oscillator \cite{Cohen2006}. Their potential application in ultra-high-density optical sweeping and switching devices, bio-mechanical and chemical sensors or electromagnetic transmitters has been mentioned \cite{ke-05-book}. Furthermore, it was shown by Jiang \textit{et al.} \cite{Jiang2002} and Zhang \textit{et al.} \cite{Zhang2004} that multifunctional nanoyarns have been fabricated by twisting multi-walled CNTs (MWCNTs) together.

Understanding the torsional behavior of CNTs for these promising applications is a fundamental issue. In recent experimental studies, Williams \textit{et al.} \cite{Williams2002} measured the torsional constants of MWCNTs using atomic force microscopy force distance technique and found that the MWCNTs become stiffer with repeated deflection. Clauss \textit{et al.} \cite{Clauss1998} presented atomically resolved scanning tunneling microscopy images of twisted armchair single-walled CNTs (SWCNTs) in a crystalline nanotube rope. Papadakis \textit{et al.} \cite{Papadakis2004} characterized nanoresonators incorporating one MWCNT as a torsional spring, and found that inter-shell mechanical coupling varies significantly from one tube to another. 

The quantum conductance of CNTs depends strongly on their atomic structure, which can be changed by twisting \cite{Cohen2006}. The change of tube's electronic properties due to twist has been predicted in several theoretical studies \cite{Rochefort1999,Liu2004}. Recently, metal-semiconducting periodic transitions were reported in experiments \cite{Cohen2006}. Moreover, Ertekin \textit{et al.} \cite{Ertekin2005} studied the ideal torsional strength and stiffness of zigzag CNTs using first-principle calculations and found that the strength of a MWCNT is about 20 times larger than that of an iron rod of the same size. Wang \textit{et al.} \cite{Wang2004} calculated the shear modulus of CNTs using molecular dynamics (MD). The mechanical integrity of SWCNTs was evaluated by Shibutani \textit{et al.} \cite{Shibutani2004} with MD simulations. In this paper, we report on the MD simulations computing the ultimate twist angle of CNTs at room temperature. Related change in the deformation energy is also investigated. The outline of this paper is as follows. The details about our computational model will be presented in Section II. The results will be shown and discussed in Section III. Then, we draw conclusions in Section IV. Analytical formulas useful for the interatomic force calculation using the AIREBO potential are given in Appendix.

\section{Methods}
To simulate the twisting of CNTs, we start with tubes fixed at one end by a hypothetical substrate and relaxed in vacuum using a Nos\'{e}-Hoover thermostat to reach equilibrium at 298 K. An imposed twist angle is then applied at the other end by successive steps of 0.1 degree every 1000 fs. The positions of atoms are updated at each iteration step (1fs) by using the leap-frog algorithm. In the AIREBO potential \cite{Stuart2000a,zhaowang-07-02,zhaowang-07-03}, the total potential energy $U^{p}$ of the system is the collection of that of individual atoms:

\begin{equation}
\label{eq:1}
U^{p}=\frac{1}{2}\sum\limits_i {\sum\limits_{j\ne i} 
{\left[ 
\begin{array}{l}
V^R(r_{ij})-b_{ij} V^A(r_{ij} ) + V_{ij}^{L-J}(r_{ij}) 
 + \sum\limits_{k\ne i,j} 
{\sum\limits_{\ell\ne i,j,k} {V_{kij\ell}^{tor} } }
\end{array}
 \right]} }
\end{equation}

where $V^R$ and $V^A$ are the interatomic repulsion and attraction terms between valence electrons, for bound atoms $i$ and $j$ at a distance $r_{ij}$. The bond order function $b_{ij}$ provides the many body effect depending on the local atomic environment of atoms $i$ and $j$. It is the key quantity which allows including the influence of the atomic environment of the bond (Huckel electronic structure theory). The long-range interactions are included by adding a parameterized \textit{Lennard-Jones} 12-6 potential term $V^{L-J}$ . $V^{tor}$ presents the torsional interactions and depends on atomic dihedral angles. Note that the long-range van der Waals interactions between atoms in the same tube must be considered in the case of large deformation, to avoid an artificial cut-off energy barrier, as discussed in Ref. \cite{Rafii-Tabar2004}. $b_{ij}$ can be written as follows.

\begin{equation}
\label{eq:2}
b_{ij} = \frac{1}{2} \left( b_{ij}^{\sigma-\pi} + b_{ji}^{\sigma-\pi} + b_{ji}^{RC}+ b_{ji}^{DH} \right)
\end{equation}

where $b_{ij}^{\sigma-\pi}$ depends on the local coordination of $i$ and $j$, and the bond angles, $b_{ji}^{RC}$ represents the influence of possible radical character of atom $j$ and of the $\pi$ bond conjugations on the bond energy. $b_{ji}^{DH}$ depends on the dihedral angle for C-C double bonds. Note that the value of $b_{ij}$ is larger for a stronger bond.

\begin{equation}
\label{eq:3}
b_{ij}^{\sigma-\pi} = \left[  
\begin{array}{l}
1+\sum\limits_{k(\neq i, j)} f^{c}_{ik}(\bm{r}_{ik}) \times G(\cos{\theta_{ijk}}) \exp(\lambda_{ijk}) + P_{ij}
\end{array}
\right]^{-1/2}
\end{equation}

where $\theta_{ijk}$ is defined as the angle between the vectors $\bm{r}_{ij}$ and $\bm{r}_{ik}$. $ P_{ij}$ and $ G(\cos{\theta_{ijk}})$ are a cubic and a fifth-order polynomial splines, respectively. The inter-atomic force is then calculated as the negative gradient of the total potential energy of the system. The formulation are presented in Appendix.

\section{Results and Discussions}
\label{rd}
\begin{figure}[ht]
\centerline{\includegraphics[width=10cm]{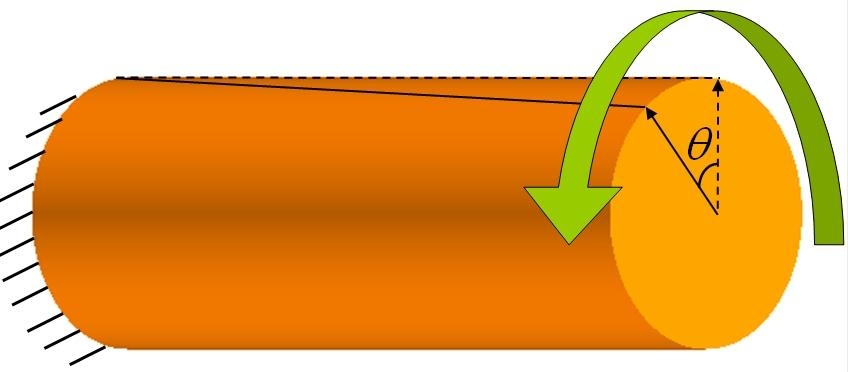}}
\caption{\label{fig:schema}
(Color online) Schematic of the definition of the twisting angle $\theta$. Imposed deformations are applied to one of the tube ends while another one is assumed to be fixed on a support.
}
\end{figure}

In this paper we studied the twist of various SWCNTs and of MWCNTs made of monochiral carbon layers, in which the interlayer distance is taken to be about 0.34 nm. The twisting angle $\theta$ is the angle between the initial position of the outer wall and its deformed position, after an imposed rotation angle at the free end, as shown in Fig. \ref{fig:schema}. 

\begin{figure}[ht]
\centerline{\includegraphics[width=14cm]{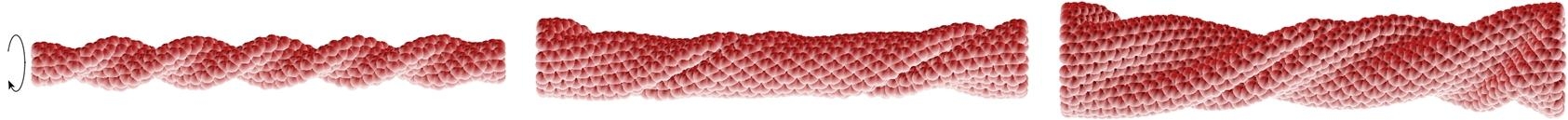}}
\caption{\label{fig:twist}
(Color online) Shape of three twisted chiral CNTs with a same length $L = 9.6$nm and a same chiral angle $= 19.1^{\circ}$, prior to fracture at $\theta= 630^{\circ}$, 497$^{\circ}$ and 427$^{\circ}$, respectively. Left: (6, 3), $R=0.31$nm; middle: (14, 7), $R=0.72$nm; right: (20, 10), $R=1.03$nm. 
}
\end{figure}

We next consider the surface change during the twisting. We observed that periodic buckling waves appear on the tube surface when a tube is largely twisted. The change of the helical shape of the CNT surface depends on the tube radius. Fig. \ref{fig:twist}. shows the different shapes of three twisted chiral CNTs prior to fracture. We can see that the buckling period is longer for big tubes than for the small one. Furthermore, we find that the length of each buckling period depends on the twisting angle and the tube radius. In our simulation, the time step between each imposed deformation is taken to be long enough (10000 step/degree) for letting the tubes have enough time to adapt to the new deformation at one end. We note that, if we apply the deformation with a higher rate (e.g. some degrees per \textit{ps}), the fracture would occur earlier and the buckling shape of the surface could be different. 

\begin{figure}[ht]
\centerline{\includegraphics[width=14cm]{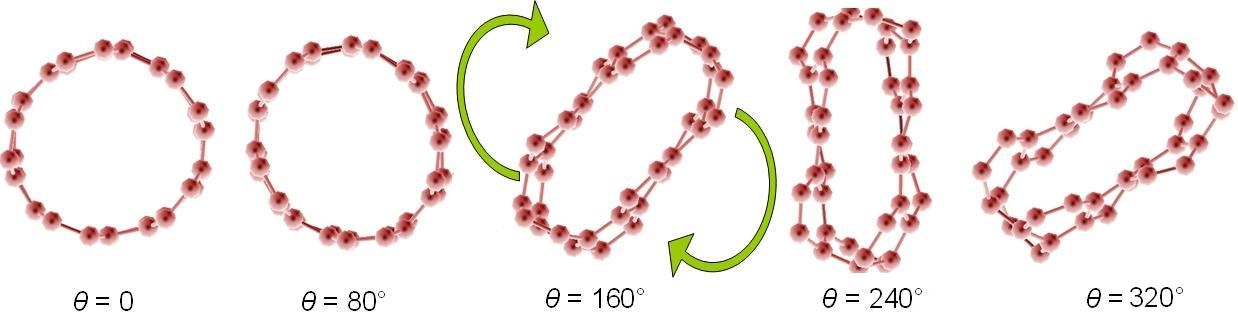}}
\caption{\label{fig:cross}
(Color online) Cross section in the middle of a (5, 5) tube ($L=9.5$nm) being twisted. The green arrows denote the direction of rotation.
}
\end{figure}

Considering that the surface twisting can significantly change the electronic properties of the tube \cite{Torrens2004}, we present in Fig. \ref{fig:cross} different shapes of the cross section of a tube twisted to several twist angles. It can be seen that the section remains circular when the deformation is relatively small. However, it deforms to an ellipse when the deformation becomes important. Then, with increasing twisting angle, this ellipse section rotates around the tube axis with an angular momentum following the direction of deformation applied to the tube end.  

\begin{figure}[ht]
\centerline{\includegraphics[width=14cm]{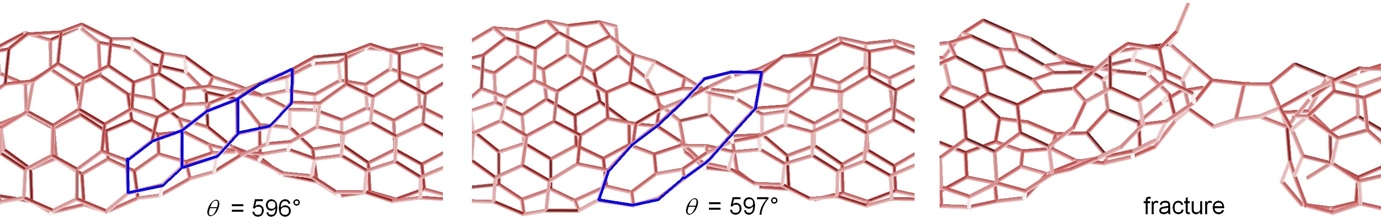}}
\caption{\label{fig:fracture}
(Color online) Fracture of a (5, 5) tube ($L=9.5$nm) being twisted to fracture.
}
\end{figure}

How much twist deformation can a CNT sustain? To answer this question, In Fig. \ref{fig:fracture}, we show the fracture of a twisted SWCNT. We can see that when $\theta = 596^\circ$, the honeycomb lattice of the tube is strongly deformed. The fracture of the tube occurs very soon (some \textit{ps}) after the appearance of the first defect. 

\begin{figure}[ht]
\centerline{\includegraphics[width=10cm]{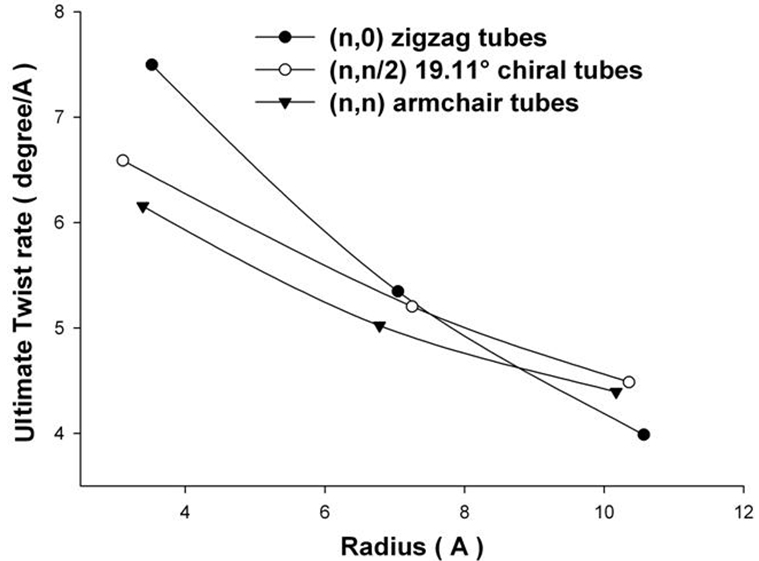}}
\caption{\label{fig:UTR}
Ultimate twist ratio versus the tube radius for 3 groups of tubes with different chiralities. Each group has 3 tubes with different radii. The length of all these tubes is fixed to 95 \AA.
}
\end{figure}

In order to present general results from the here-studied short tubes, we define the twist ratio as the twisting angle $\theta$ per unit length of CNTs. We plot in Fig. \ref{fig:UTR} the ultimate value of the twist ratio (UTR) for 9 SWCNTs of a same length but with different radii and chiralities. It can be seen that the UTR of the small tubes is clearly higher than that of the big ones. We can also see that the UTR of the zigzag tubes decreases faster than that of the armchair tubes with increasing tube radius. We can conclude that a big tube can resist better to twist than a small one. 

\begin{figure}[ht]
\centerline{\includegraphics[width=9cm]{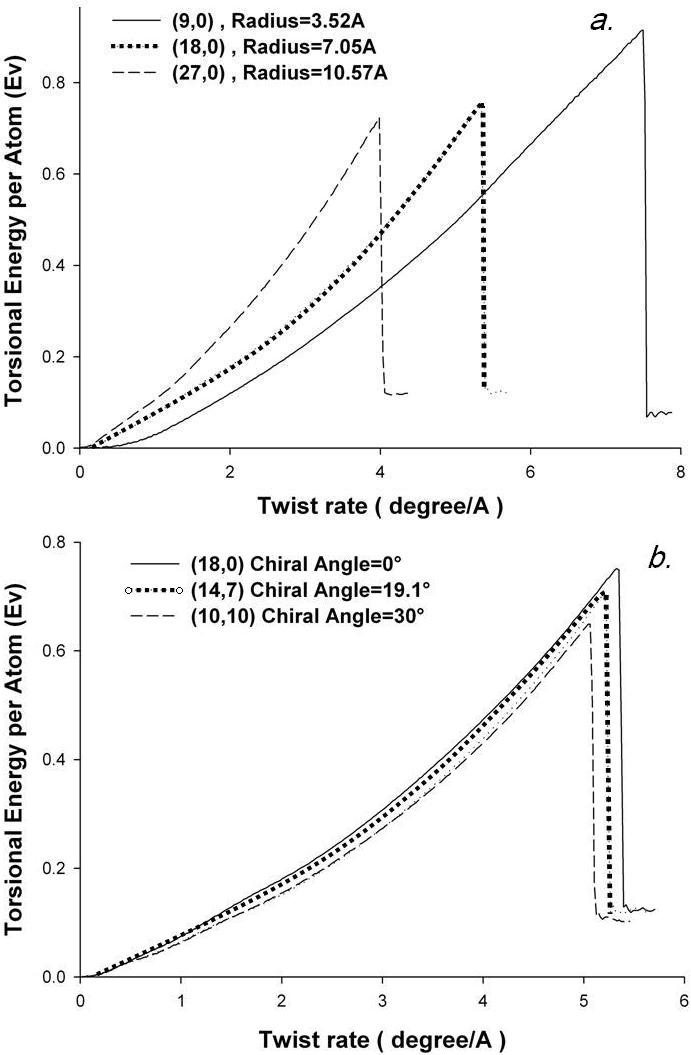}}
\caption{\label{fig:Ener}
Deformation energy versus the twist ratio for: (a) 3 zigzag tubes with the same length but with different radii, and (b) 3 tubes with almost the same length and radius but with different chiral angles. The deformation energy presented here is the value averages on all the atoms.
}
\end{figure}

To show further effect of the tube geometry, we define the deformation energy of the tube as the change of the total potential energy of the CNT. It is an important factor coupled to the tube's elastic constant. We plot in Fig. \ref{fig:Ener} the torsion energy as a function of the twist ratio. We can see from Fig. \ref{fig:Ener} (a) that the deformation energy of a big tube increases faster than that of a small one, while the ultimate value of the deformation energy for a big tube is lower than that for a small one. In Fig. \ref{fig:Ener} (b). we use the tubes of similar radii and lengths to show that the deformation energy is almost independent of the tube chirality. The increase ratio of deformation energy of the zigzag tube is slightly higher than that of the chiral and the armchair ones. This corresponds to the fact that the average axial bond strength of a zigzag SWCNT is slightly higher than that of other tubes with similar sizes but differing chiralities \cite{Van2000,Wang2009}.

\begin{figure}[ht]
\centerline{\includegraphics[width=13cm]{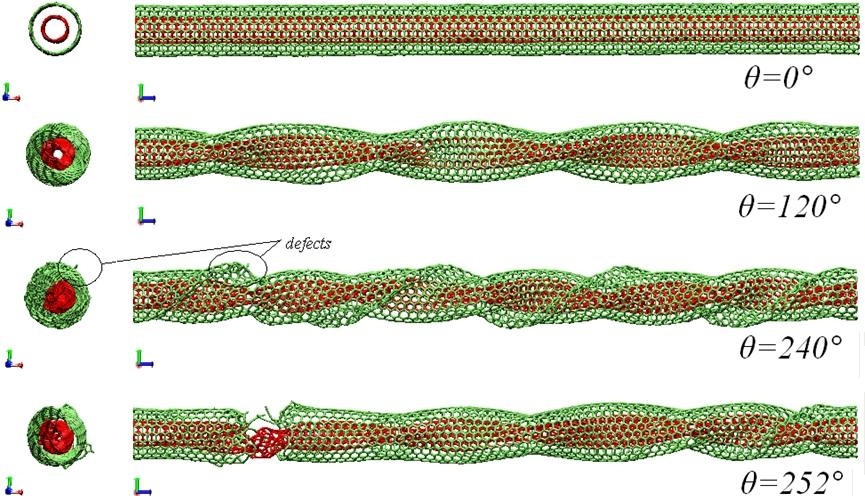}}
\caption{\label{fig:MWCNT}
(Color online) Fracture of an MWCNT (5,5)@(10,10) ($L=194.3$\AA , $R =(3.39$\AA@$6.78$\AA))
}
\end{figure}

We study also the twist of multi-walled CNTs (MWCNTs), as demonstratiod in Fig. \ref{fig:MWCNT}. It shows from two positions of observation how an armchair MWCNT breaks under twist. We can see the appearance of buckling waves in both the inner and outer layers when the tube is deformed, while the fracture occurs first at the outer layer after the appearance of defects on its surface. 

\begin{figure}[ht]
\centerline{\includegraphics[width=10cm]{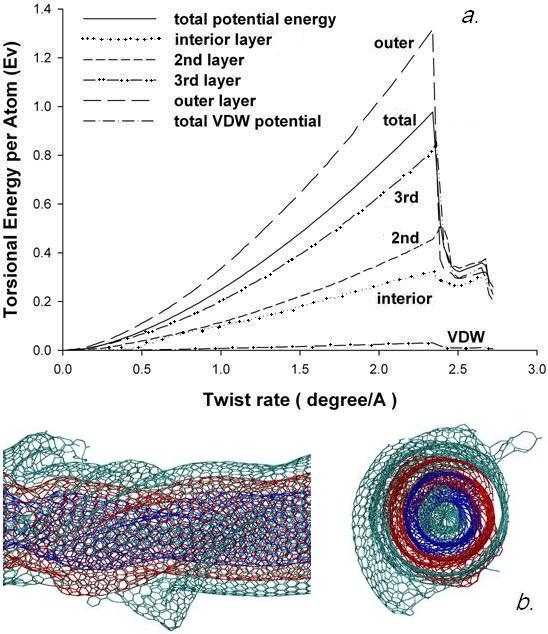}}
\caption{\label{fig:MWCNT2}
(Color online) (a) Average deformation energy (per atom) vs. twist ratio for each layer in a MWCNT (0,9)@(0,18)@ (0,27)@(0,36) $L=84.0$\AA. The deformation energy is the average value per atom. (b) Failure of this MWCNT being twisted. Top. side view. Bottom. cross-section view.
}
\end{figure}

We show the ultimate twist ratio of MWCNTs in Table 1. It can be seen that the UTR decreases with the number of carbon layers. The deformation energy of a zigzag MWCNT is plotted in Fig. \ref{fig:MWCNT2} (a). We can see that the energy per atom of the outer layer during the deformation is much higher than that of the inner layer. This can explain why the tubes are always broken from the outer layer when they are largely twisted. We can also see that the the van der Waals interaction does not play a very important role in the total deformation energy. In Fig. \ref{fig:MWCNT2} (b). we can see the corresponding image of the failure of the twisted tube.

\section{Conclusions}
In summary, the twist of CNTs has been simulated by using the MD method based on the AIREBO potential. Surface transition from zigzag or armchair to chiral type and periodic buckling waves were observed in our simulations. We also observed the creation of defects and the fracture on the tube surface.  The cross section of SWCNTs is found to become an elliptic and rotates around the tube center axis when the deformation is large enough. We calculated the ultimate value of the twist ratio and the deformation energy for several types of CNTs with different geometries. We find that the small tubes can be twisted more than the big ones. The ultimate twist ratio of zigzag MWCNTs is higher than armchair ones. Moreover, analytical formulas useful for the interatomic force calculation using the AIREBO potential are given in Appendix. 

\newpage

\section*{Acknowledgments} 
We gratefully thank S. J. Stuart for the numerics. This work was done as parts of the CNRS GDR-E Nb 2756. Z. W. acknowledges the support received from the region of Franche-Comt\'{e} (grant 060914-10). 

\newpage

\section*{Figures}

\newpage

\newpage

\newpage

\newpage

\newpage

\newpage

\newpage

\newpage

\section*{Table}

\begin{table}[tb]
\begin{center}
\begin{tabular}{c|c}
MWCNTs & UTR(degree/\AA) \\
\hline
(5,5)@(10,10) & 2.99 \\
(5,5)@(10,10)@ (15,15) @ (20,20) & 1.64 \\
(5,5)@(10,10)@ ... @ 30,30)      & 0.96\\
(0,9)@(0,18) & 3.66 \\
(0,9)@(0,18)@ (0,27)@(0,36) & 2.36 \\
(0,9)@(0,18)@ ... @(0,54)   & 1.55 \\
\end{tabular}
\end{center}
\caption{\label{table:MWCNTS} Ultimate twist ratio of MWCNTs with the same length about 200 \AA.}
\end{table}

\newpage

\bibliographystyle{unsrt}

\end{document}